\RequirePackage{ifpdf}
\ifpdf 
\documentclass[pdftex]{sigma}
\else
\documentclass{sigma}
\fi

\begin{document}
\allowdisplaybreaks

\renewcommand{\PaperNumber}{021}

\FirstPageHeading

\ShortArticleName{Degenerate Multiplicity of the $sl_2$ Loop
Algebra for the 6V Model at Roots of Unity}

\ArticleName{On the Degenerate Multiplicity of the
$\boldsymbol{sl_2}$ Loop\\ Algebra for the 6V Transfer Matrix at
Roots of Unity}

\Author{Tetsuo DEGUCHI}

\AuthorNameForHeading{T. Deguchi}

\Address{Department of Physics, Faculty of Science,  Ochanomizu University, \\
2-1-1 Ohtsuka, Bunkyo-Ku, Tokyo 112-8610, Japan}
\Email{\href{mailto:deguchi@phys.ocha.ac.jp}{deguchi@phys.ocha.ac.jp}}

\ArticleDates{Received October 31, 2005, in f\/inal form February
06, 2006; Published online February 17, 2006}

\Abstract{We review the main result of cond-mat/0503564. The
Hamiltonian of the XXZ spin chain and the transfer matrix of the
six-vertex model has the $sl_2$ loop algebra symmetry if the $q$
parameter is given by a root of unity, $q_0^{2N}=1$, for an
integer $N$. We discuss the dimensions of the degenerate
eigenspace generated by a regular Bethe state in some sectors,
rigorously as follows:
 We show that every regular Bethe ansatz eigenvector in the sectors is
a highest weight vector and derive the highest weight
 ${\bar d}_k^{\pm}$,
 which leads to evaluation parameters~$a_j$.
If the evaluation parameters are distinct, we obtain the
dimensions of the highest weight representation generated by the
regular Bethe state.}


\Classification{81R10; 81R12; 81R50; 81V70}

\section{Introduction}

The XXZ spin chain is one of the most important exactly solvable
quantum systems. The Hamiltonian under the periodic boundary
conditions is given by
\begin{gather}
H_{XXZ} =  {\frac 1 2} \sum_{j=1}^{L} \left(\sigma_j^X
\sigma_{j+1}^X +
 \sigma_j^Y \sigma_{j+1}^Y + \Delta \sigma_j^Z \sigma_{j+1}^Z  \right) .
\label{hxxz}
\end{gather}
Here the XXZ anisotropy $\Delta$ is related to the $q$ parameter
by  $\Delta= (q+q^{-1})/2$. The XXZ Hamiltonian (\ref{hxxz}) is
derived from the logarithmic derivative of the transfer matrix of
the six-vertex model, and hence they have the same set of
eigenvectors.

 Recently, it was explicitly shown that when $q$ is a root of unity
 the XXZ Hamiltonian commutes with the generators of
the $sl_2$ loop algebra~\cite{DFM}. Let $q_0$ be a primitive root
of unity satisfying $q_0^{2N}=1$ for an integer $N$. We introduce
operators $S^{\pm(N)}$ as follows
\begin{gather*}
S^{\pm(N)}= \sum_{1 \le j_1 < \cdots < j_N \le L} q_0^{{N \over 2
} \sigma^Z} \otimes \cdots \otimes q_0^{{N \over 2} \sigma^Z}
\otimes \sigma_{j_1}^{\pm} \otimes q_0^{{(N-2) \over 2} \sigma^Z}
\otimes  \cdots \otimes q_0^{{(N-2) \over 2} \sigma^Z}
\nonumber \\
\phantom{S^{\pm(N)}= }{} \otimes \sigma_{j_2}^{\pm} \otimes
q_0^{{(N-4) \over 2} \sigma^Z} \otimes \cdots \otimes
\sigma^{\pm}_{j_N} \otimes q_0^{-{N \over 2} \sigma^Z} \otimes
\cdots \otimes q_0^{-{N \over 2} \sigma^Z}   .
\end{gather*}

They are derived from the $N$th power of the generators $S^{\pm}$
of the quantum group $U_q(sl_2)$ or~$U_q(\hat{sl_2})$. We also
def\/ine  $T^{(\pm)}$ by the complex conjugates of $S^{\pm(N)}$,
i.e.\ $T^{\pm (N)} = \left( S^{\pm (N)} \right)^{*}$. The
operators,  $S^{\pm(N)}$ and $T^{\pm (N)}$, generate the $sl_2$
loop algebra, $U(L(sl_2))$, in the sector
\begin{gather}
S^Z \equiv 0 \quad ({\rm mod} \, N)  . \label{sct0}
\end{gather}
 Here the value of the total spin
$S^Z$ is given by an integral multiple of $N$. In the sector
(\ref{sct0}),  the ope\-rators $S^{\pm (N)}$ and $T^{\pm (N)}$
(anti-)commute with the  transfer matrix of the six-vertex
model~$\tau_{6V}(v)$,  and they commute with the Hamiltonian of
the XXZ spin chain \cite{DFM}:
\begin{gather*}
{[}S^{\pm(N)},H_{XXZ} {]}={[}T^{\pm(N)},H_{XXZ} {]}=0  .
\end{gather*}

One of the most important physical questions is to obtain the
degenerate multiplicity of the $sl_2$ loop algebra. Let us denote
by  $| R \rangle$ a regular Bethe state with $R$ down spins. Here
we def\/ine regular Bethe states by such Bethe ansatz eigenvectors
that are constructed from f\/inite and distinct solutions of the
Bethe ansatz equations, whose precise def\/inition will be given
in Section~2.
 For any given regular Bethe state in the sector (\ref{sct0}),
we may have the following degenerate eigenvectors of the XXZ
Hamiltonian
\[
S^{- (N)} | R \rangle , \quad T^{- (N)} | R \rangle  , \quad
\big(S^{-(N)}\big)^2 | R \rangle  ,  \quad T^{- (N)} S^{+(N)} T^{- (N)} |
R \rangle  , \quad \ldots .
\]
However, it is nontrivial how many of them are linearly
independent. The number should explain the degree of the spectral
degeneracy. We thus want to know the dimensions of the degenerate
eigenspace generated by the Bethe state $| R \rangle$.  Here we
note that some of the spectral degeneracies of the XYZ spin chain
(the eight-vertex model) at roots of unity were f\/irst discussed
by Baxter~\cite{Baxter123}
 (see also \cite{Baxter02,Baxter04}). In fact, there are
such spectral degeneracies of the XYZ spin chain at roots of unity
that are closely related to the $sl_2$ loop algebra symmetry of
the XXZ spin chain~\cite{Missing,CSOS,FM-8vertex,FM-8vertex2}.
Quite interestingly, various aspects of the spectral degeneracies
of the XXZ spin chain have been discussed by several authors from
dif\/ferent viewpoints
\cite{Alcaraz,Korepanov94,Korepanov95,Pasquier,Tarasov-cyclic}.

Recently, there has been some progress on the degenerate
multiplicity as far as regular Bethe states in some sectors such
as (\ref{sct0}) are concerned~\cite{HWT}.
 For the XXZ spin chain at roots of unity,
 Fabricius and McCoy had made important observations
 on degenerate multiplicities of the $sl_2$ loop algebra,
and conjectured the `Drinfeld polynomials of Bethe ansatz
eigenvectors' \cite{FM1,FM2,Odyssey}. However, it was not clear
whether the representation generated by a Bethe state is
irreducible or not. Here we remark that a Drinfeld polynomial is
def\/ined for an irreducible f\/inite-dimensional representation.
Motivated by the previous results \cite{FM1,FM2,Odyssey}, an
algorithm for determining the dimensions of the representation
generated by a given regular Bethe state has been rigorously
formulated in some sectors such as (\ref{sct0}) and some
restricted cases of evaluation parameters $a_j$, which will be
def\/ined in Section~5. We show rigorously \cite{HWT} that every
regular Bethe state in the sectors is highest weight, assuming a
conjecture that the Bethe roots are continuous with respect to the
$q$ parameter at roots of unity. We evaluate the highest 
weight~${\bar d}_k^{\pm}$ for the Bethe state, which leads to evaluation
parameters $a_j$. Here  ${\bar d}_k^{\pm}$ will be def\/ined in
Section~4. Independently, it has been shown \cite{Chari-P3} that
if the evaluation parameters of a highest weight representation of
the $sl_2$ loop algebra are distinct, then it is irreducible.
 It thus follows that the `Drinfeld polynomial
 of a~regular Bethe state' corresponds to
the standard Drinfeld polynomial def\/ined for the irreducible
representation generated by the regular Bethe state, if the
evaluation parameters are distinct.
 The conjecture of Fabricius and McCoy has been proved at least
 in the sectors such as (\ref{sct0}) and in the case of
 distinct evaluation parameters.
  Thus, the purpose of the paper is to review
the main points of the complete formulation of~\cite{HWT}
brief\/ly with some illustrative examples. Some other aspects of
the $sl_2$ loop algebra symmetry have also been
discussed~\cite{Andrei,Korff-McCoy,Tarasov}.

The contents of the paper is given as follows. In Section~2, we
introduce Bethe ansatz equations and regular solutions of the
Bethe ansatz equations. We def\/ine regular Bethe states. In
Section~3, we discuss brief\/ly the $sl_2$ loop algebra symmetry
of the XXZ spin chain at roots of unity. Here, the conditions of
roots of unity are specif\/ied precisely. In Section~4, we explain
the Drinfeld realization of the $sl_2$ loop algebra, i.e.\ the
classical analogues of the Drinfeld realization of the quantum
af\/f\/ine group $U_q({\hat{sl}}_2)$.  In Section~5, we review the
algorithm for determining the degenerate multiplicity of a regular
Bethe state in the sectors. Here we note that Theorem~\ref{th:reg}
generalizes the $su(2)$ symmetry of the XXX spin chain shown by
Takhtajan and Faddeev~\cite{TF}. In Section~6 we discuss some
examples of the Drinfeld polynomials explicitly.

\section{Bethe ansatz equations and the transfer matrix}
\subsection{Regular solutions of the Bethe ansatz equations}

Let us assume that a set of complex numbers,
 ${\tilde t}_1, {\tilde t}_2, \ldots, {\tilde t}_R$ satisfy
the Bethe ansatz equations at a root of unity:
\begin{gather}
\left( {\frac {\sinh({\tilde t}_j + \eta_0)}
 {\sinh({\tilde t}_j - \eta_0)}} \right)^L
= \prod_{k=1; k \ne j}^{M} {\frac {\sinh({\tilde t}_j - {\tilde
t}_k + 2 \eta_0)} {\sinh({\tilde t}_j - {\tilde t}_k - 2 \eta_0)}}
\, , \qquad \mbox{for} \quad j=1, 2, \ldots, R. \label{BAEatRTU}
\end{gather}
Here the parameter $\eta$ is def\/ined by the relation
$q=\exp(2\eta)$, and $\eta_0$ is given by $q_0=\exp(2 \eta_0)$. If
a given set of solutions of the Bethe ansatz equations are
f\/inite and distinct, we call them {\it regular}. We call a set
of solutions of the Bethe ansatz equations {\it Bethe roots}.

A set of regular solutions of the Bethe ansatz equations leads to
an eigenvector of the XXZ Hamiltonian. We call it a {\it regular
Bethe state} of the XXZ spin chain or a {\it regular XXZ Bethe
state}, brief\/ly. Explicit expressions of regular Bethe states
are derived through the algebraic Bethe ansatz
method~\cite{Korepin}.

\subsection{Transfer matrix of the six-vertex model}

We now introduce $L$ operators for the XXZ spin chain. Let $V_n$
be two-dimensional vector spaces for $n=0, 1, \ldots, L$. We
def\/ine an operator-valued matrix $L_n(z)$ by
\begin{gather*}
 L_n(z)
= \left(
\begin{array}{cc}
L_n(z)^1_1  &  L_n(z)^1_2 \\
 L_n(z)^2_1  & L_n(z)^2_2
\end{array}
\right) =  \left(
\begin{array}{cc}
\sinh \left( z \, I_n + \eta \sigma_n^z \right)
& \sinh 2 \eta \, \sigma_n^{-} \\
\sinh 2 \eta \, \sigma_n^{+} & \sinh \left( z \, I_n - \eta
\sigma_n^z \right)
\end{array}
\right)  .
\end{gather*}
Here $L_n(z)$ is a matrix acting on the auxiliary vector space
$V_0$, where $I_n$ and $\sigma_n^a$ ($a=z, \pm$) are operators
acting on the $n$th vector space $V_n$. The symbol $I$ denotes the
two-by-two identity matrix,
 $\sigma^{\pm}$ denote $\sigma^{+}= E_{12}$ and $\sigma^{-} = E_{21}$,
and $\sigma^x$, $\sigma^y$, $\sigma^z$ are the Pauli matrices.

We def\/ine the monodromy matrix $T$ by the product:
\begin{gather*}
T(z) = L_L(z) \cdots L_2(z) L_1(z)  .
\end{gather*}
Here the matrix elements of $T(z)$  are given by
\begin{gather*}
T(z) = \left( \begin{array}{cc}
A(z) & B(z) \\
C(z) & D(z)
\end{array}
\right).
\end{gather*}

We def\/ine the transfer matrix of the six vertex model
$\tau_{6V}(z)$ by the following trace:
\begin{gather*}
\tau_{6V}(z) = {\rm tr} \, T(z) = A(z) + D(z).
\end{gather*}
We call the transfer matrix {\it homogeneous}. It is invariant
under lattice translation.

\section[The $sl_2$ loop algebra symmetry at roots of unity]{The $\boldsymbol{sl_2}$
loop algebra symmetry at roots of unity}

We shall show the $sl_2$ loop algebra symmetry of the XXZ spin
chain at roots of unity in some sectors. We shall discuss two
cases with even $L$ and odd $L$.

\subsection{Roots of unity conditions}

Let us explicitly formulate roots of unity conditions as follows.
\begin{definition}[Roots of unity conditions]
We say that $q_0$ is a root of unity with $q_0^{2N}=1$, if one of
the three conditions hold: (i) $q_0$ is a primitive $N$th root of
unity with $N$ odd ($q_0^N=1$); (ii) $q_0$ is a primitive $2N$th
root of unity with $N$ odd ($q_0^N=-1$); (iii) $q_0$ is a
primitive $2N$th root of unity with $N$ even ($q_0^N=-1$). We call
the cases (i) and (iii)  type I, the case (ii)  type II.
\label{df:roots}
\end{definition}
In the case of  $S^Z \equiv 0$ (mod $N$) we consider all the three
conditions of roots of unity.  However, in the case of $S^Z \equiv
N/2$ (mod $N$) with $N$ odd, we consider only the condition (i) of
roots of unity, i.e.\
 $q_0$ is a primitive $N$th root of unity with $N$ odd ($q_0^N=1$).

\subsection{(Anti-)commutation relations at roots of unity}

We show the $sl_2$ loop algebra symmetry of the XXZ spin chain in
the following two sectors: (a) in the sector $S^{Z} \equiv 0$ (mod
$N$) where $q_0$ is a root of unity with $q_0^{2N}=1$, as
specif\/ied in Def\/inition~\ref{df:roots}; (b) in the sector
$S^{Z} \equiv N/2$ (mod $N$) with $N$ odd where $q_0$ is a
primitive $N$th root of unity.

Let us assume that there exists a set of regular solutions of
Bethe ansatz equations (\ref{BAEatRTU}) with $R$ down-spins,
i.e.\ $R$ regular Bethe roots. We also assume that the lattice
size $L$, the number of regular Bethe roots $R$ and the integer
$N$ satisfy the following relation:
\begin{gather}
L - 2 R = n N  , \qquad n \in {\mathbb Z}. \label{LRn}
\end{gather}
If $n$ is even, the regular Bethe state $|R \rangle$ is in the
sector $S^Z \equiv 0$ (mod $N$), while if $n$ is odd and $N$ is
also odd, then it is in the sector $S^Z \equiv N/2$ (mod $N$).
Here we recall that the symbol $| R \rangle$ denotes the regular
Bethe state constructed from the given $R$ regular Bethe roots.

It has been shown \cite{DFM} that operators $S^{\pm (N)}$ and
$T^{\pm (N)}$ (anti-)commute with the  transfer matrix of the
six-vertex model $\tau_{6V}(v)$ in the sector $S^{Z} \equiv 0$
(mod $N$) at
 $q_0$ with $q_0^{2N}=1$
\begin{gather}
S^{\pm (N)} \, \tau_{6V}(z)=q_0^N \tau_{6V}(z) \, S^{\pm (N)} ,
\qquad T^{\pm (N)} \, \tau_{6V}(z)=q_0^N \tau_{6V}(z) \, T^{\pm
(N)}   . \label{loop}
\end{gather}
Furthermore, it is also shown \cite{HWT} that in the sector $S^{Z}
\equiv N/2$ (mod $N$) when $N$ is odd and  $q_0$ satisf\/ies
$q_0^N=1$, operators $S^{\pm (N)}$ and $T^{\pm (N)}$
(anti-)commute with the  transfer matrix of the six-vertex model
$\tau_{6V}(v)$.

 From the (anti-)commutation relations (\ref{loop}) it follows
that the operators $S^{\pm (N)}$  and $T^{\pm (N)}$ commute with
the XXZ Hamiltonian in the sector $S^Z \equiv 0$ (mod $N$) when
$q_0$ satisf\/ies $q_0^{2N}=1$, and in the sector $S^Z \equiv N/2$
when $N$ is odd and $q_0^N=1$. Here we recall that the XXZ
Hamiltonian $H_{XXZ}$ is given by the logarithmic derivative of
the (homogeneous) transfer matrix $\tau_{6V}(v)$.

\subsection[The algebra generated by $S^{\pm (N)}$ and $T^{\pm (N)}$]{The algebra
generated by $\boldsymbol{S^{\pm (N)}}$ and $\boldsymbol{T^{\pm
(N)}}$}

Let us discuss the algebra generated by
 the operators \cite{DFM}, $S^{\pm (N)}$ and $T^{\pm (N)}$.
 When  $q_0$ is of type~I,
we have the following identif\/ication~\cite{DFM}:
\begin{gather}
E_0^{+}=T^{-(N)}, \qquad E_0^{-}=T^{+(N)}, \qquad
E_1^{+}=S^{+(N)}, \qquad
E_1^{-}=S^{-(N)}, \nonumber\\
- H_0 =  H_1 = {\frac 2 N}  S^Z  . \label{id1}
\end{gather}
When $q$ is of type II, we have the following \cite{Missing}:
\begin{gather}
E_0^{+} = \sqrt{-1} \, T^{-(N)}, \qquad  E_0^{-} =  \sqrt{-1} \,
T^{+(N)}, \qquad
 E_1^{+}=  \sqrt{-1} \,  S^{+(N)}, \nonumber\\
E_1^{-} =  \sqrt{-1} \,  S^{-(N)},
 \qquad - H_0 = H_1=  {\frac 2 N}  S^Z  .
\label{imaginary}
\end{gather}
Here $\sqrt{-1}$ denotes the square root of $-1$ (cf.\ (A.13) of
\cite{Missing}; see also \cite{Korff-McCoy}). The operators
$E_j^{\pm}$ and~$H_j$ for $j=0, 1$, are the Chevalley generators
of the af\/f\/ine Lie algebra $\hat{sl}_2$. In fact, the operators
$E_j^{\pm}$, $H_j$ for $j=0,1$, satisfy the def\/ining relations
\cite{Kac} of the $sl_2$ loop algebra \cite{DFM}:
\begin{gather}
  H_0 + H_1  = 0  , \qquad
{\rm [} H_i, E_j^{\pm} {\rm ]}  = \pm a_{ij} E_j^{\pm}  ,
  \qquad i,j = 0, 1,
 \label{Cartan} \\
  {\rm [} E_i^{+}, E_j^{-} {\rm ]}  =  \delta_{ij} H_{j}  ,
  \qquad i,j = 0, 1, \label{EF} \\
 {\rm [} E_i^{\pm},  {\rm [} E_i^{\pm}, {\rm [} E_i^{\pm},
 E_j^{\pm} {\rm ]}   {\rm ]}   {\rm ]} = 0 ,
\qquad i,j=0,1, \quad i \ne j. \label{Serre}
\end{gather}
Here, the Cartan matrix $(a_{ij})$ of $A_1^{(1)}$ is def\/ined  by
\begin{gather*}
\left(
\begin{array}{cc}
a_{00} & a_{01} \\
a_{10} & a_{11}
\end{array}
\right)
 = \left(
\begin{array}{cc}
2 & -2 \\
-2 & 2
\end{array}
\right).
\end{gather*}
The Serre relations (\ref{Serre}) hold if $q_0$ is a primitive
$2N$th root of unity, or a primitive $N$th root of unity with $N$
odd~\cite{DFM}. We derive it through the higher order quantum
Serre relations due to Lusztig~\cite{Lusztig}. The Cartan
relations (\ref{Cartan}) hold for generic $q$. The relation
(\ref{EF}) holds for the identif\/ication (\ref{id1}) when $q_0$
is a~root of unity of type~I, and for the identif\/ication
(\ref{imaginary}) when $q_0$ is a~root if unity of type II.

In the sector $S^Z \equiv 0$ (mod $N$)  we have
 the commutation relation \cite{DFM}:
\begin{gather*}
\big[ S^{+(N)}, S^{-(N)}\big] = (-1)^{N-1} q^{N} \, {\frac 2 N}
S^Z .
\end{gather*}
Here the sign factor $(-1)^{N-1} q^{N}$ is given by  1 or $-1$
 when $q$ is a root of unity of type I or II, respectively.
In the case of the sector $S^Z \equiv N/2$ (mod $N$) with $N$ odd
and $q_0$ a primitive $N$th root of unity,  we have
 the following commutation relation:
\begin{gather*}
\big[ S^{+(N)}, S^{-(N)}\big] =  {\frac 2 N} S^Z .
\end{gather*}

\subsection{Some remarks on quantum groups at roots of unity}
 Let the symbol $U_{q}^{\rm res}(g)$ denote
the algebra generated by the $q$-divided powers of the Chevalley
generators of a Lie algebra $g$ \cite{Chari-P2}. The
correspondence of the algebra $U_{q_0}^{\rm res}(g)$ at a root of
unity, $q_0$, to the Lie algebra $U(g)$ was obtained essentially
through the machinery introduced by Lusztig~\cite{Modular,Lusztig}
both for f\/inite-dimensional simple Lie algeb\-ras and
inf\/inite-dimensional af\/f\/ine Lie algeb\-ras. In fact, by
using the higher order quantum Serre relations~\cite{Lusztig}, it
has been shown that the af\/f\/ine Lie algebra $U(\hat{sl}_2)$ is
generated by the operators such as $S^{\pm(N)}$ at  roots of
unity. However, in the case of the af\/f\/ine Lie algebras
$\hat{g}$, the highest weight conditions for the Drinfeld
generators are dif\/ferent from those for the Chevalley
generators. Through the highest weight vectors of the Drinfeld
generators, f\/inite-dimensional representations were discussed by
Chari and Pressley for $U_{q_0}^{\rm res}(\hat{g})$
\cite{Chari-P2}.

\section[The Drinfeld realization of the $sl_2$ loop algebra]{The Drinfeld realization of the $\boldsymbol{sl_2}$
loop algebra}

{\samepage Finite-dimensional representations of the $sl_2$ loop
algebra,  $U(L(sl_2))$, are derived by taking the classical
analogues of the Drinfeld realization of the quantum $sl_2$ loop
algebra, $U_q(L(sl_2))$~\cite{Chari-P1,Chari-P2}. The classical
analogues of the Drinfeld generators, ${\bar x}_k^{\pm}$ and
${\bar h}_k$ ($k \in {\mathbb Z}$),
 satisfy the def\/ining relations in the following:
\begin{gather*}
[\bar{h}_j, \bar{x}_{k}^{\pm} ] = \pm 2 \bar{x}_{j+k}^{\pm} ,
\qquad [\bar{x}_j^{+}, \bar{x}_k^{-} ] = \bar{h}_{j+k}  , \qquad
{\rm for} \  j, k \in {\mathbb Z}  .
\end{gather*}
Here $[{\bar h}_j, {\bar h}_{k} ]=0$ and $[{\bar x}_j^{\pm}, {\bar
x}_k^{\pm}] =0$ for $j, k \in {\mathbb Z}$.}

Let us now def\/ine highest weight vectors. In a representation of
$U(L(sl_2))$, a vector $\Omega$ is called {\it a highest weight
vector} if  $\Omega$ is annihilated by generators ${\bar
x}_{k}^{+}$ for all integers $k$ and such that $\Omega$ is a
simultaneous eigenvector of every generator of the Cartan
subalgebra, ${\bar h}_k$ ($k\in {\mathbb
Z}$)~\cite{Chari-P1,Chari-P2}:
\begin{gather}
{\bar x}_k^{+} \Omega =  0  , \qquad {\rm for} \  k \in {\mathbb
Z} ,
\label{eq:annihilation} \\
{\bar h}_{k} \Omega  =  {\bar d}_k^{+} \Omega  , \qquad {\bar
h}_{-k} \Omega = {\bar d}_{-k}^{-} \Omega  , \qquad {\rm for} \  k
\in {\mathbb Z}_{\ge 0} . \label{eq:Cartan}
\end{gather}
We call a representation of $U(L(sl_2))$ {\it highest weight} if
it is generated by a highest weight vector. The set of the complex
numbers ${\bar d}_k^{\pm}$ given in (\ref{eq:Cartan}) is called
{\it the highest weight}. It is shown \cite{Chari-P1} that every
f\/inite-dimensional irreducible representation is highest weight.
To a f\/inite-dimensional irreducible representation $V$ we
associate a unique polynomial through the highest weight ${\bar
d}_k^{\pm}$~\cite{Chari-P1}. We call it the Drinfeld polynomial.
Here the degree $r$ is given by the weight ${\bar d}_0^{\pm}$.

It is easy to see that the highest weight vector of a
f\/inite-dimensional irreducible representation~$V$ is a
simultaneous eigenvector of operators $({\bar x}_{0}^{+})^k ({\bar
x}_{1}^{-})^k/(k!)^2$ for  $k>0$, and the Drinfeld polynomial of
the representation $V$ has another  expression as
follows~\cite{HWT}
\begin{gather}
P(u) =\sum_{k=0}^{r} \lambda_k (-u)^k  , \label{DrinfeldP}
\end{gather}
where  $\lambda_k$ denote  the eigenvalues of operators $({\bar
x}_{0}^{+})^k ({\bar x}_{1}^{-})^k/(k!)^2$. It is noted that the
author learned the expression of the Drinfeld
polynomial~(\ref{DrinfeldP}) from Jimbo~\cite{Jimbo-summer} (see
also \cite{FM1,Odyssey}).

\section{Algorithm for evaluating the degenerate multiplicity}
\subsection[A useful theorem on the $sl_2$ loop algebra]{A useful theorem on the $\boldsymbol{sl_2}$ loop algebra}

Let $\Omega$ be a highest weight vector and $V$ the representation
generated by $\Omega$. Here
 $V$ is not necessarily irreducible.
Suppose that $V$ is f\/inite-dimensional and ${\bar h}_0 \Omega =
r \, \Omega$. We def\/ine a~polynomial~$P_{\Omega}(u)$ by the
relation (\ref{DrinfeldP}) with $\lambda_k$. We show that the
roots of the polynomial $P_{\Omega}(u)$ are nonzero and f\/inite,
and the degree of  $P_{\Omega}(u)$ is given by $r$~\cite{HWT}. Let
us factorize $P_{\Omega}(u)$ as
\begin{gather*}
P_{\Omega}(u) = \prod_{k=1}^{s} (1 - a_k u)^{m_k} ,
\end{gather*}
where $a_1, a_2, \ldots, a_s$ are distinct, and their
 multiplicities are given by  $m_1, m_2, \ldots, m_s$, respectively.
Then, we call  $a_j$ the {\it evaluation parameters} of $\Omega$.
Here we note that $r$ is given by the sum: $r=m_1 + \cdots + m_s$.
If all the multiplicities are given by 1, i.e.\ $m_j=1$ for
$j=1,2, \ldots, s$, we say that evaluation parameters $a_1, a_2,
\ldots, a_r$ are distinct. In the case of distinct evaluation
parameters (i.e.\ $m_j=1$ for all $j$), we have the following
theorem \cite{Chari-P3}:

\begin{theorem}
Every finite-dimensional highest weight representation of the
$sl_2$ loop algebra with distinct evaluation parameters $a_1, a_2,
\ldots, a_r$ is irreducible. Furthermore, it has dimensions $2^r$.
\label{th:irrep}
\end{theorem}

Theorem~\ref{th:irrep} plays an important role in connecting the
polynomial $P_{\Omega}(u)$ with an irreducible
f\/inite-dimensional representation.  In fact, if $V$ has distinct
evaluation parameters, then it is irreducible, and the polynomial
$P_{\Omega}(u)$ is equivalent to the Drinfeld polynomial of $V$.
Moreover, we rigorously  obtain the dimensions of $V$.

Theorem \ref{th:irrep} has been shown in~\cite{Chari-P3} by using
the fact that the specialized irreducible modules for the quantum
algebra are quotients of the Weyl module. Here we note that
theorem \ref{th:irrep} is also derived by constructing explicitly
a basis of  a  highest weight representation with distinct
evaluation parameters~\cite{distinct}.

\subsection{Regular Bethe states as highest weight vectors}

 For the XXZ spin chain at roots of unity,
 Fabricius and McCoy made important observations
 on the highest weight conjecture  \cite{FM1,FM2,Odyssey}.
Motivated by them, we discuss the following:
\begin{theorem}
{\rm (i)} Every regular Bethe state $|R \rangle$ in the sector
$S^Z \equiv 0$ $({\rm mod}\, N)$ at $q_0$ is a highest weight vector of
the $sl_2$ loop algebra. Here $q_0$ is a root of unity with
$q_0^{2N}=1$, as specified in Definition~{\rm \ref{df:roots}}.
{\rm (ii)} Every regular Bethe state $|R \rangle$ in the sector
$S^Z \equiv N/2$ $({\rm mod}\,N)$ at $q_0$ is a~highest weight vector of
the $sl_2$ loop algebra. Here $N$ is odd and  $q_0$ is a primitive
$N$th root of unity. \label{th:reg}
\end{theorem}

The Theorem~\ref{th:reg} is proved in~\cite{HWT} by assuming the
conjecture that  for a given regular Bethe state in the sector
$S^Z \equiv 0$ (mod\,$N$) (or $S^Z \equiv N/2$ (mod\,$N$))
 the set of solutions of the Bethe ansatz equations
 are continuous with respect to
the parameter $q$ at the root of unity~$q_0$. In some cases such
as $R=0$ or~1, the conjecture is trivial.

By the method of the algebraic Bethe ansatz, we  derive the
following relations~\cite{HWT}:
\begin{gather}
S^{+(N)} \, | R \rangle  =  T^{+(N)} \, | R \rangle
 =  0  , \nonumber\\
\big(S^{+(N)} \big)^k \big(T^{-(N)} \big)^{k}/(k!)^2 \, | R
\rangle  = {\cal Z}_k^{+} \, | R \rangle \qquad \mbox{for}
\  k \in {\mathbb Z}_{\ge 0} , \nonumber \\
 \big(T^{+(N)} \big)^k \big( S^{-(N)} \big)^{k}/(k!)^2 \,
 | R \rangle  =  {\cal Z}_k^{-} \, | R \rangle \qquad
\mbox{for} \  k \in {\mathbb Z}_{\ge 0}  . \label{rel}
\end{gather}
Here, the operators $S^{\pm(N)}$, $T^{+(N)}$,  $T^{-(N)}$ and
$2S^Z/N$ satisfy the same def\/ining relations of the $sl_2$ loop
algebra as
 generators ${\bar x}_{0}^{\pm}$, ${\bar x}_{-1}^{+}$, ${\bar x}_1^{-}$ and
${\bar h}_0$, respectively, and hence the relations (\ref{rel})
correspond to~(\ref{eq:annihilation}) and~(\ref{eq:Cartan}).

In equations (\ref{rel}) eigenvalues ${\cal Z}_k^{\pm}$ are
explicitly evaluated as follows
\begin{gather*}
{\cal Z}_k^{+} =(-1)^{kN} {\tilde{\chi}}_{kN}^{+}  , \qquad {\cal
Z}_k^{-} =(-1)^{kN} {\tilde{\chi}}_{kN}^{-}   .
\end{gather*}
Here the ${\tilde{\chi}}_m^{\pm}$ have been def\/ined by the
coef\/f\/icients of the following expansion with respect to small
$x$:
\begin{gather*}
{\frac {\phi(x)} {{\tilde F}^{\pm} (xq_0) {\tilde
F}^{\pm}(xq_0^{-1})} } =
 \sum_{j=0}^{\infty} {\tilde{\chi}}_j^{\pm} \, x^j ,
\end{gather*}
where $\phi(x)=(1-x)^L$ and ${\tilde
F}^{\pm}(x)=\prod\limits_{j=1}^{R} (1 - x \exp(\pm 2 {\tilde
t}_j))$.

\subsection{Drinfeld polynomials of regular Bethe states\\ and
the degenerate multiplicity}

Let  $|R\rangle$ be a regular Bethe state at $q_0$ in one of the
sectors specif\/ied in Theorem~\ref{th:reg}. The Drinfeld
polynomial of the regular Bethe state  $|R\rangle$ is explicitly
derived by putting $\lambda_k=(-1)^{kN} {\tilde{\chi}}_{kN}^{+} $
into equation~(\ref{DrinfeldP}). Here, the coef\/f\/icients
$\tilde{\chi}_{kN}^{\pm}$ are explicitly evaluated as \cite{HWT}
\begin{gather}
{\tilde{\chi}}_{kN}^{\pm} = \sum_{n=0}^{{\rm min}(L, kN)} (-1)^n
 \left(
\begin{array}{c}
L \\
n
\end{array}
\right)  \sum_{n_1 + \cdots + n_R =kN-n} e^{\pm
\sum\limits_{j=1}^{R} 2 n_j \tilde{t}_j }  \prod_{j=1}^{R} [n_j +
1]_{q_0} . \label{eq:chi-kN}
\end{gather}
Here $[n]_q=(q^n-q^{-n})/(q-q^{-1})$ and the sum is taken over all
nonnegative integers $n_1, n_2, \ldots, n_R$ satisfying $n_1 +
\cdots + n_R =kN-n$: when $R=0$, $n$ is given by $n=kN$.

For the case of distinct evaluation parameters, we obtain the
algorithm for the degeneracy of a regular Bethe state as follows.
\begin{corollary}
Let $|R \rangle$ be a regular Bethe state such as specified in
Theorem~{\rm \ref{th:reg}}. If the Drinfeld polynomial of the
representation $V$ generated by $| R \rangle$ gives
 evaluation parameters $a_j$ with multiplicities $m_j=1$
 for $j=1, 2, \ldots, r$,
then we have $\dim  V = 2^{r}$,  where $r=(L-2R)/N$.
\end{corollary}

\section{Examples of Drinfeld polynomials of Bethe states}

\subsection[The vacuum state with even $L$]{The vacuum state with even $\boldsymbol{L}$}

We now calculate the Drinfeld polynomial $P(u)$ for the the vacuum
state $| 0 \rangle$ where $L=6$ and $N=3$ with $q_0^3=1$. When $N$
is odd and $q_0^N =1$, we have $\lambda_k^{+} =  {\cal Z}_k^{+} =
(-1)^{k} {\tilde \chi}_{kN}^{+} $.
 From the formula (\ref{eq:chi-kN}) we have
\begin{gather*}
{\tilde{\chi}}_3^{+} = (-1)^3  \left(
\begin{array}{c}
6 \\
3
\end{array}
 \right)
= - 20
 , \qquad
{\tilde{\chi}}_6^{+} = (-1)^6  \left(
\begin{array}{c}
6 \\
6
\end{array}
 \right) = 1  .
\end{gather*}
Thus we have $\lambda_{1}^{+}= 20$ and $\lambda_2^{+}= 1$, and the
Drinfeld polynomial is given by
\begin{gather*}
P(u) = 1 - 20 u +  u^2  .
\end{gather*}
Here, the evaluation parameters are given by
\begin{gather*}
a_1, a_2 =  10 \pm 3 {\sqrt{11}}  .
\end{gather*}
We note that the two evaluation parameters are distinct, and the
degree of $P$ is two, i.e.\ $r=2$, $m_1=m_2=1$. Therefore, the
degenerate multiplicity is given by $2^2=4$.

\subsection[The vacuum state with odd $L$]{The vacuum state with odd $\boldsymbol{L}$}

Let us calculate the Drinfeld polynomial $P(u)$ for the odd $L$
case. We consider the the vacuum state $| 0 \rangle$ where $L=9$
and $N=3$ with $q_0^3=1$. The vacuum state $|0 \rangle$ is in the
sector $S^Z \equiv 3/2$ (mod\,$3$), since $S^Z = 9/2 = 3/2 + 3$.

 From the formula (\ref{eq:chi-kN}) we have
\begin{gather*}
{\tilde{\chi}}_3^{+} = (-1)^3  \left(
\begin{array}{c}
9 \\
3
\end{array}
 \right)
= - 84
 , \qquad
{\tilde{\chi}}_6^{+} = (-1)^6  \left(
\begin{array}{c}
9 \\
6
\end{array}
 \right) = 84  , \qquad
{\tilde{\chi}}_9^{+} = (-1)^9  \left(
\begin{array}{c}
9 \\
9
\end{array}
 \right) = -1  .
\end{gather*}
Thus we have $\lambda_{1}^{+}= 84$, $\lambda_2^{+}= 84$, and
$\lambda_{3}^{+}= 1$. The Drinfeld polynomial is given by
\begin{gather*}
P(u) = 1 - 84 u +  84 u^2 - u^3  .
\end{gather*}
Here, the evaluation parameters are given by
\begin{gather*}
 a_1, a_2 = {\frac 1 2 } \left( 83 \pm 9 {\sqrt{85}}  \right).
\end{gather*}
We note that the three evaluation parameters are distinct, and the
degree of $P$ is three, i.e.\ $r=3$, $m_1=m_2=m_3=1$. Therefore,
the degenerate multiplicity is given by $2^3=8$.

\subsection[The regular Bethe state with one down-spin ($R=1$)]{The regular Bethe state with one
down-spin ($\boldsymbol{R=1}$)}

{\samepage
 For the case of $R=1$, the Bethe ansatz equations at generic $q$ are given by
\begin{gather*}
\left({\frac {\sinh(t_j + \eta)}
             {\sinh(t_j - \eta)} }
\right)^L = 1 \qquad {\rm for} \  j =0, 1, \ldots, L-1 .
\end{gather*}
 We solve the Bethe ansatz equations in terms of variable $\exp(2t_j)$
as follows
\begin{gather}
\exp(2t_j) ={ \frac {1- \omega_j q} {q- \omega_j}} \qquad {\rm
for} \  j=0, 1, \ldots, L-1, \label{eq:sol}
\end{gather}
where $\omega_j$ denotes an $L$th root of unity: $\omega_j =
\exp\left(2 \pi \sqrt{-1} j/L \right)$, for $j=0, 1, \ldots,
L-1$.}

Let us assume that the regular Bethe state with one down-spin is
in the sector $S^Z \equiv 0$ (mod\,$N$) and $q_0$ be a root of
unity with $q_0^{2N}=1$, or in the sector $S^Z \equiv N/2$
(mod\,$N$) where
 $q_0$ is a primitive $N$th root of unity with $N$ odd.
Then we have from (\ref{eq:chi-kN})
\begin{gather*}
 {\tilde{\chi}}^{+}_{kN}  =  \sum_{\ell=0}^{\min (nN+2, kN)}
(-1)^j [kN+1-\ell]_{q_0} \left(
\begin{array}{c}
nN + 2 \\
\ell
\end{array}
\right)
 \left(  { \frac {1- \omega_j  q_0} {q_0 - \omega_j}}  \right)^{kN - \ell} .
\end{gather*}
Here we have from (\ref{LRn}) that $L=nN +2$ when $R=1$.

Let us consider the case of $N=3$, and $q_0= \exp(\pm 2 \pi
\sqrt{-1}/3 )$. Here $L=8$. The regular Bethe state with rapidity
$\tilde{t}_2$ (the case of $j=2$ in equation~(\ref{eq:sol})) has
the Drinfeld polynomial in the following:
\begin{gather*}
P(u)= 1 - 13 \left(2-\sqrt{3}\right) u + \left(7 - 4
\sqrt{3}\right) u^2,
\end{gather*}
where the evaluation parameters $a_1$ and $a_2$ are given by
\begin{gather*}
a_1  ,  a_2 = {\frac 1 2} \left(13 \pm \sqrt{165}\right) \left(2
-\sqrt{3}\right).
\end{gather*}
The dimensions of the highest weight representation generated by
the Bethe state are therefore given by $2^2=4$.

\subsection*{Acknowledgements}

The author would like to thank K.~Fabricius and B.M.~McCoy for
useful discussions. He would also like to thank the organizers for
their kind invitation to the Sixth International Conference
``Symmetry in Nonlinear Mathematical Physics'', June 20--26, 2005,
Institute of Mathematics, Kyiv, Ukraine. He is grateful to many
participants of the conference for useful comments and
discussions. This work is partially supported by the Grant-in-Aid
(No.~17540351).

\LastPageEnding


\begin{thebibliography}{99}
\footnotesize

\bibitem{Alcaraz} Alcaraz F.C., Grimm U., Rittenberg V.,
The XXZ Heisenberg chain, conformal invariance and the operator
content of $c<1$ systems, {\it  Nucl. Phys. B}, 1989, V.316,
735--768.


\bibitem{Baxter123} Baxter R.J.,
Eight-vertex model in lattice statistics and one-dimensional
anisotropic Heisenberg chain. I.~Some fundamental eigenvectors,
{\it Ann. Phys.}, 1973, V.76, 1--24;
 II. Equivalence to a generalized Ice-type lattice model,
{\it Ann. Phys.}, 1973, V.76, 25--47; III. Eigenvectors of the
transfer matrix and Hamiltonian, {\it Ann. Phys.}, 1973, V.76,
48--71.


\bibitem{Baxter02}  Baxter R.J.,
Completeness of the Bethe ansatz for the six and eight vertex
models, {\it J. Statist. Phys.}, 2002, V.108, 1--48;
cond-mat/0111188.


\bibitem{Baxter04}  Baxter R.J.,
The six and eight-vertex models revisited, {\it J. Statist.
Phys.}, 2004, V.116, 43--66; cond-mat/0403138.

\bibitem{Andrei} Braak D., Andrei N.,
 On the spectrum of the XXZ-chain at roots of unity,
{\it J. Statist. Phys.}, 2001, V.105, 677--709; cond-mat/0106593.


\bibitem{Chari-P1} Chari V., Pressley A.,
Quantum af\/f\/ine algebras, {\it Comm. Math. Phys.}, 1991, V.142,
261--283.

\bibitem{Chari-P2} Chari V., Pressley A.,
Quantum af\/f\/ine algebras at roots of unity, {\it Represent.
Theory}, 1997, V.1, 280--328; q-alg/9609031.

\bibitem{Chari-P3} Chari V., Pressley A., Weyl modules for classical and
 quantum af\/f\/ine algebras, {\it Represent. Theory}, 2001, V.5, 191--223; math.QA/0004174.


\bibitem{Missing} Deguchi T.,
Construction of some missing eigenvectors of the XYZ spin chain at
the discrete coupling constants and the exponentially large
spectral degeneracy of the transfer matrix, {\it J. Phys. A: Math.
Gen.}, 2002, V.35, 879--895; cond-mat/0109078.


\bibitem{CSOS} Deguchi T.,
 The 8V CSOS model and the $sl_2$ loop algebra symmetry
 of the six-vertex model at roots of unity,
{\it Internat. J. Modern Phys. B}, 2002, V.16, 1899--1905;
cond-mat/0110121.


\bibitem{HWT} Deguchi T.,
 XXZ Bethe states as highest weight vectors
of the $sl_2$ loop algebra at roots of unity, cond-mat/0503564.

\bibitem{distinct} Deguchi T.,
The six-vertex model at roots of unity and some highest weight
representations of the $sl_2$ loop algebra, in preparation (to be
submitted to the Proceedings of RAQIS'05, Annecy, France).


\bibitem{DFM} Deguchi T., Fabricius K.,  McCoy B.M.,
The $sl_2$ loop algebra symmetry of the six-vertex model at roots
of unity, {\it J. Statist. Phys.}, 2001, V.102, 701--736;
cond-mat/9912141.

\bibitem{FM1} Fabricius K., McCoy B.M.,
Bethe's equation is incomplete for the XXZ model at roots of
unity, {\it J. Statist. Phys.}, 2001, V.103, 647--678;
cond-mat/0009279.

\bibitem{FM2} Fabricius K., McCoy B.M.,
Completing Bethe's equations at roots of unity, {\it J. Statist.
Phys.}, 2001, V.104, 573--587; cond-mat/0012501.

\bibitem{Odyssey} Fabricius K., McCoy B.M.,
Evaluation parameters and Bethe roots for the six-vertex model at
roots of unity, {\it Progress in Mathematical Physics}, Vol.~23
 (MathPhys Odyssey 2001), Editors  M.~Kashiwara and T.~Miwa, Boston, Birkh{\"a}user, 2002, 119--144; cond-mat/0108057.


\bibitem{FM-8vertex} Fabricius K., McCoy B.M.,
New developments in the eight-vertex model,
{\it J. Statist Phys.}, 2003, V.111, 323--337; cond-mat/0207177.\\
Fabricius K., McCoy B.M., Functional equations and fusion matrices
for the eight-vertex model, {\it Publ. Res. Inst. Math. Sci.},
2004, V.40, 905--932; cond-mat/0311122.

\bibitem{FM-8vertex2} Fabricius K., McCoy B.M.,
New developments in the eight-vertex model II. Chains of odd
length, cond-mat/0410113.

\bibitem{Jimbo-summer} Jimbo M., Private communication, July 2004.

\bibitem{Kac} Kac V., Inf\/inite dimensional Lie algebras,  Cambridge,
Cambridge University Press, 1990.


\bibitem{Korepanov94} Korepanov I.G., Hidden symmetries in the 6-vertex model
of statistical physics, {\it Zap. Nauchn. Sem. \mbox{S.-Peter\-burg.}
Otdel. Mat. Inst. Steklov. (POMI)}, 1994, V.215, 163--177 (English
transl.: {\it J. Math. Sci. (New York)}, 1997, V.85, 1661--1670);
  hep-th/9410066.

\bibitem{Korepanov95} Korepanov I.G., Vacuum curves of the ${\cal L}$-operators related to the six-vertex model,
{\it St. Petersburg Math. J.}, 1995, V.6, 349--364.



\bibitem{Korepin} Korepin V.E., Bogoliubov N.M., Izergin A.G.,
Quantum inverse scattering method and correlation functions,
Cambridge, Cambridge University Press,  1993.


\bibitem{Korff-McCoy} Korf\/f C., McCoy B.M.,
Loop symmetry of integrable vertex models at roots of unity, {\it
Nucl. Phys.~B}, 2001, V.618, 551--569; hep-th/0104120.



\bibitem{Modular} Lusztig G., Modular representations and
quantum groups, {\it Contemp. Math.}, 1989, V.82, 59--77.

\bibitem{Lusztig} Lusztig G., Introduction to quantum groups, Boston,
Birkh{\"a}user, 1993.


\bibitem{Pasquier} Pasquier V., Saleur H.,
Common structures between f\/inite systems and conformal f\/ield
theories through quantum groups, {\it Nucl. Phys. B}, 1990, V.330,
523--556.



\bibitem{TF} Takhtajan L., Faddeev L.,
Spectrum and scattering of excitations in the one-dimensional
isotropic Heisenberg model, {\it J. Sov. Math.}, 1984, V.24,
241--267.


\bibitem{Tarasov-cyclic} Tarasov V.O.,
 Cyclic monodromy matrices for the $R$-matrix of the six-vertex model
and the chiral Potts model with f\/ixed spin boundary conditions,
in Inf\/inite Analysis, Part A, B (Kyoto, 1991), {\it Adv. Ser.
Math. Phys.}, Vol.~16, River Edge, NJ, World Sci. Publishing,
1992, 963--975.



\bibitem{Tarasov} Tarasov V.O., On the Bethe vectors
for the XXZ model at roots of unity, math.QA/0306032.


\end{thebibliography}
\end{document}